\documentclass[%
 reprint,
 nofootinbib, 
 amsmath,amssymb,
 aps,
]{revtex4-2}

\usepackage{graphicx}
\usepackage{dcolumn}
\usepackage{bm}
\usepackage{hyperref}
\usepackage{gensymb}
\usepackage{subfigure}
\usepackage{float}
\usepackage{xcolor}

\begin{document}

\preprint{ArXiv:}

\title{Characterizing Beam Profiles in Accelerator Neutrino Experiments through Off-Axis Neutrino Interactions}

\author{S. Karpova}
\email{Svetlana.Karpova@unige.ch}
\author{F. S\'anchez}%
\author{D. Douqa}%

\affiliation{Universit\'e de Gen\`eve, Section de Physique, DPNC, 1205 Gen\`eve, Switzerland}%

\date{\today}

\begin{abstract}
We introduce a novel approach that utilizes neutrino events from the off-axis near detector to investigate the beam profile in long-baseline neutrino experiments. Understanding the dynamics of the neutrino beam is crucial for improving the precision of neutrino oscillation measurements. We demonstrate that certain observables related to the azimuthal angle of the neutrino direction are useful for determining the average neutrino production point from experimental data, providing a valuable cross-check against Monte Carlo simulations. Additionally, these observables can help identify potential alignment issues between the detector and the decay volume. In future neutrino experiments with significantly higher statistics, these observables will become essential to ensure the accuracy and stability of the beam profile.
\end{abstract}

\maketitle


\section{Introduction}
\label{sec:intro}

There are several long-baseline neutrino experiments currently operating worldwide that study neutrino properties by observing the effects of neutrino oscillations over long distances.  
A precise understanding of the neutrino beam production, propagation, and dynamics is essential for improving the accuracy and reliability of oscillation measurements.  
Typically, beam dynamics are characterized by a combination of the neutrino flux and interaction cross sections. Despite the complexity of reconstructing neutrino interactions, the leading source of systematic uncertainty in cross-section measurements is the flux $ \phi_f (E) $. Beam diagnostics remain one of the few tools available to substantially reduce flux uncertainties ~\cite{app11041644}.

In conventional accelerator-based neutrino beams, neutrinos are primarily produced when a high-energy proton beam strikes a fixed target, generating charged mesons such as pions ($\pi$) and kaons ($K$). Subsequently, these mesons decay into neutrinos, predominantly muon neutrinos ($ \nu_\mu $) and anti-muon neutrinos ($ \bar{\nu}_\mu $), with a small contamination of electron neutrinos ($ \nu_e $) ~\cite{kopp2007accelerator}. Therefore, $\pi$ and $K$ are considered the parent particles of these neutrinos. The difference in the masses of these two particles lead to significantly different kinematics of the emitted neutrinos, an effect that becomes even more pronounced in off-axis beam configurations. This distinction can be exploited to evaluate their relative contributions.

The most effective approach to studying neutrino oscillations involves the combination of near and far detectors. A near detector, positioned close to the neutrino source (where oscillations are negligible), characterizes the initial neutrino flux and spectrum. In contrast, a far detector, located at a distance near the oscillation peak, observes the effects of the neutrino oscillation ~\cite{app11041644}. Near detectors play a crucial role in constraining the properties of the neutrino beam, such as its flux and energy spectrum, thus reducing systematic uncertainties. To achieve a narrower neutrino energy spectrum, detectors are often placed off-axis relative to the neutrino beam, at a specific angle with respect to the beamline ~\cite{beavis1995long}. The ND280 detector of the T2K experiment is an example of such an off-axis detector, used in neutrino oscillation experiments, with an off-axis angle of 2.5$^{\circ}$ ~\cite{ferrero2009nd280}. 

The decay volume, where the parent particles decay into neutrinos, typically extends several hundred meters, with the parent particles decaying at different positions along this path.

In current neutrino experiments, the standard approach for determining the average decay position of parent particles relies on Monte Carlo (MC) simulations, with this average position subsequently applied to experimental data. Since the neutrino production point cannot be directly observed in the experimental data, this method is essential. However, it has several limitations:  

\begin{itemize}  
    \item The decay volume is relatively long, leading to variations in the off-axis angle along its length rather than aligning with a single average value. This variation influences the observed neutrino energy spectrum.

    \item Although the average decay position varies with the energy and type of parent particles, experimental analyses often approximate it using a fixed value, simplifying the underlying physical dependencies.

    \item Misalignments in the beam-detector configuration or inaccuracies in the simulation can result in discrepancies between the average decay position inferred from MC simulations and that in the experimental data, potentially exposing systematic issues.
\end{itemize}
Given these challenges, it is crucial to evaluate the agreement between MC predictions and experimental results. To address this, we introduce a novel approach to studying the neutrino beam, where the influence of cross-section and flux is significantly reduced. By leveraging a new observable that is minimally dependent on the dynamics of neutrino interactions at first order, we can directly compute the average decay position of parent particles as a function of neutrino energy. More importantly, this technique can be applied to both simulations and experimental data through event reconstruction analysis.

In this paper, we focus on low-energy Charged Current Quasi-Elastic (CCQE) events to illustrate the dependencies observed in a scenario similar to T2K ~\cite{abe2011t2k}. Specifically, we study the reconstruction of CC0$\pi$1p events with a near off-axis detector. This interaction channel is not only one of the most dominant at low neutrino energies but also one of the best understood in this energy range.
The remainder of this paper is organized as follows: Section~\ref{sec:beam} describes neutrino interactions and beam modeling. Section~\ref{sec:observable} introduces the experimental observable. Section~\ref{sec:results} discusses the information that can be extracted from this observable. Conclusions are drawn in Section~\ref{sec:conclusions}.  

\section{Beam modeling and neutrino interaction simulations}
\label{sec:beam}

\subsection{Neutrino interaction Monte Carlo Model}

We use the NEUT event generator ~\cite{hayato2021neut} to simulate neutrino interactions with carbon, the primary target material for the T2K experiment. NEUT provides accurate predictions for neutrino interactions across different channels, including CCQE interactions, which dominate at T2K energies, as well as Charged Current Resonant (CCRes), Deep Inelastic Scattering (DIS), and Shallow Inelastic Scattering (SIS). Additionally, interactions involving two nucleons producing two holes, known as 2p2h events, are also relevant.  
 
To model the initial nuclear state for CCQE interactions, NEUT employs several nuclear models, including the Relativistic Mean Field, Local Fermi Gas, and Spectral Function approaches. Furthermore, NEUT simulates the transport of final-state hadrons as they exit the nucleus using an intranuclear cascade model (INC) to account for Final State Interactions (FSI), which modifies the hadronic composition of the interaction.  
Among all the events generated by NEUT, we select the CC0$\pi$1p event topology, which includes a muon ($\mu$), a proton ($p$), and no mesons in the final state after nuclear rescattering. This is the typical selection strategy for highly segmented experiments, such as the near detector of T2K. These events contain a majority of CCQE events, with small fractions of CCRes, DIS and 2p2h events.  

\subsection{Event selection and reconstruction }

We generate all possible neutrino-carbon interactions modeled by NEUT, weighted according to the T2K neutrino flux ~\cite{T2K:2012bge}. Subsequently, we apply an event selection procedure closely resembling that used in the T2K experiment, such that the final state particles in each event fulfill the following conditions ~\cite{T2K:2018rnz}:
\begin{enumerate}
    \item A visible $\mu$ with momentum $>250~\text{MeV}/c$ and $\cos(\theta_\mu) > -0.6$. 
    \item No charged $\pi$ in the final hadronic system. 
    \item One and only one $p$ with momentum $>450~\text{MeV}/c$ and $\cos(\theta_p) > 0.4$.  
\end{enumerate}

In NEUT, we use the true neutrino direction directly from the simulation. Additionally, we reconstruct the neutrino direction using the true $\mu$ and $p$ momenta, given by:
\begin{equation}
\vec{P_\nu} = \vec{P_\mu} + \vec{P_p}.
\end{equation}
We refer to this reconstruction as {\it RecoTrue}.  

To reconstruct the neutrino energy, we use the following equation: 
\begin{equation}
E_\nu = E_\mu + E_p + S_{RE} - M_n,
\end{equation}
where $S_{RE}$ represents the removal energy, fixed to its average value of 28 MeV for carbon-12 ~\cite{douqa2024superscaling}.  

The events are categorized into four energy regions:  
\begin{itemize}
    \item 0.5~GeV $< E_{\nu} \le $ 0.75~GeV,  
    \item 0.75~GeV $< E_{\nu} \le $ 1~GeV,  
    \item 1~GeV $< E_{\nu} \le $ 2~GeV,  
    \item 2~GeV $ < E_{\nu}$.
\end{itemize}

\subsection{Decay volume and the detector position}

Long-baseline neutrino experiments always include a decay volume, as mentioned above, to allow parent particles to decay into neutrinos. To simulate the neutrino beam's decay volume, we adopt an exponential decay distribution, where neutrinos are predominantly produced near the target, and their production rate decreases exponentially along the decay volume. This serves as a simplified representation of the actual decay process. Our beam model is inspired by the T2K configuration, treating the decay volume as a straight section with an exponential decay profile.  For simplicity, the decay volume is approximated as a line source along the beam axis, neglecting its finite transverse dimensions. This is justified by the horn-focusing that causes most neutrino-producing decays to occur near the central axis. As a result, the transverse spread predominantly induces a spatial smearing of the effective neutrino production vertex, which remains well characterized along the beam axis. Equivalently, this can be interpreted as an effective production position corresponding to the point of closest approach between the neutrino trajectory and the beam axis. The overall effect of this approximation is a reduction in the sensitivity of the observable proposed in this letter, and it must be evaluated under each specific experimental condition. 

The decay volume originates at the target station ($\vec r_{tgt}$) with coordinates $(0., 0.333, -5.0497)$~m, extends in the direction ($\hat d_{dcy}$) given by $(0., -0.0645, 0.999)$, and has a total length of 110~m. The mean decay position varies with the neutrino energy, and since we analyze the results as a function of energy, the specific decay profile changes accordingly.  

\begin{figure}[h]
    \centering
\includegraphics[width=1.0\linewidth]{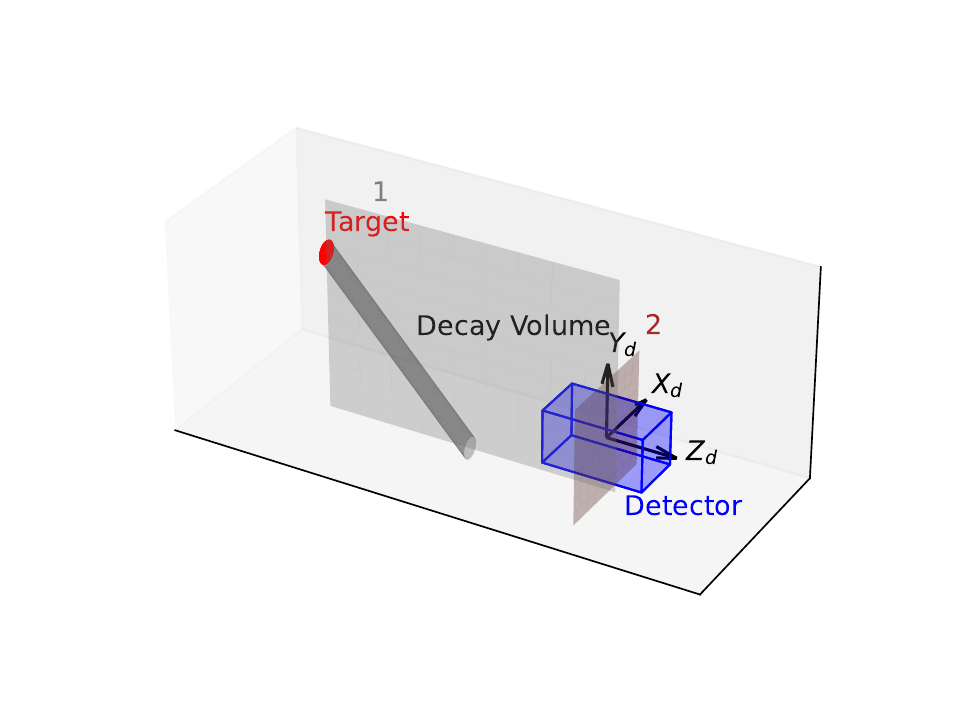}
    \caption{3D schematic diagram (not to scale) illustrating the relative positions of the target, decay volume, and detector. The gray plane labeled 1 corresponds to the target-decay volume plane (YZ plane), while the brown plane labeled 2 represents the detector plane (XY plane) with the coordinate axes inside the detector.}
    \label{fig:target_DV_detector_3D}
\end{figure}

For our study we used a near detector. The coordinates of the center of the detector are set to $(0.,-8.146,280.1)$~m following the T2K geometry, which will serve as the baseline for the current study. However, we have modified the X coordinate to force the near detector to be in the same plane as the decay volume and the target.
To illustrate the spatial configuration, a 3D schematic diagram is shown in Fig.~\ref{fig:target_DV_detector_3D}. The target, decay volume, and detector are represented, with numbered planes indicating key components. The figure highlights the spatial relationships between these elements; it is not to scale and is intended solely to clarify the conceptual layout.
In the diagram, plane 1 (gray) corresponds to the target–decay volume plane (YZ plane), while plane 2 (brown) represents the vertical detector plane (XY plane). The red disk marks the target location, and the blue cuboid depicts the detector.

\section{Neutrino direction as a beam characterization observable }
\label{sec:observable}

\begin{figure}[h]
    \centering
    \subfigure[]{
        \includegraphics{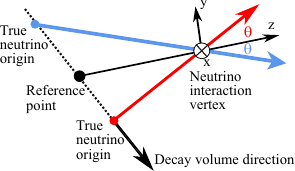}
        \label{fig:subfig1}
    }
    \vskip 1em 
    \subfigure[]{
        \includegraphics{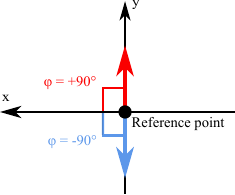}
        \label{fig:subfig2}
    }
    \caption{The polar $\theta$ (a) and azimuthal $\varphi$ (b) angles of the neutrino direction.}
    \label{fig:theta_phi}
\end{figure}

As mentioned in the Introduction, we investigate the decay position of the neutrino's parent particle, that is, the origin of the neutrino. To achieve this, we require knowledge of the neutrino direction, specifically the polar angle ($\theta$) and azimuthal angle ($\varphi$) at the interaction point with respect to the reference system (see Fig. \ref{fig:theta_phi}).

This direction encodes information on the relative alignment of the detector with the decay volume and the neutrino production point within the decay volume. The line connecting the disintegration point of the parent particle to the interaction vertex in the detector coincides with the reconstructed neutrino direction.  

The position of the detector relative to the decay volume can vary between experiments. However, it is always possible to place the detector within the plane that contains both the target and the decay volume. This configuration allows for more precise angle measurements, as all neutrinos will have the same azimuthal angles. Positioning the detector outside the beam plane would cause the measured angles to depend on the exact decay point of the neutrino within the decay volume.

When properly measured, this information, combined with the reconstructed neutrino energy, provides valuable insights into beam dynamics. $ \nu_\mu $ are primarily produced by $\pi$ and $K$ decays, but their spatial distribution within the decay volume differs given the distinct boosts, decay times, and masses of the parent particles. These differences affect the neutrino energy in the off-axis configuration. As a result, the spatial distribution of events along the decay volume may serve as an indirect measurement of the relative contributions of $\pi$ and $K$. 


\subsection{Reconstructed Neutrino direction }
\label{sec:reconstructed_neutrino_direction}
In the reconstruction of the neutrino direction, the accuracy of the reconstructed neutrino direction is degraded by the effects of the detector, but primarily by the Fermi momentum of the target nucleon in the nucleus, following the impulse approximation.

In particular, $\theta$ tends to be poorly reconstructed. In contrast, $\varphi$ statistically retains more information about the true direction of the neutrino. To exploit this property, we introduce a potential disintegration point in the decay volume, called the reference point ($\vec{r}$).  

Using the reference point ($\vec{r}$), we define a reference frame where the $X$ axis ($\vec{x}$) is perpendicular to the plane containing the decay volume direction $\vec d_{decay}$ and the neutrino interaction vertex ($\vec{v}$) at the detector:
\begin{equation}
\vec{x} = \frac{(\vec{v} - \vec{r}) \times \vec d_{decay}} {| (\vec{v} - \vec{r}) \times \vec d_{decay} | },
\end{equation}
\vspace{-0.6cm}
\begin{equation}
\vec{z} = \frac{\vec{v} - \vec{r}} {|\vec{v} - \vec{r}  | },
\end{equation}
\vspace{-0.6cm}
\begin{equation}
\vec{y} = \frac { \vec{z} \times \vec{x} }  { | \vec{z} \times \vec{x}   | }.
\end{equation}

This coordinate system provides a structured approach to analyzing neutrino directionality while mitigating the impact of Fermi motion on $\theta$.

Notably, the value of $\varphi$ rotates approximately 180$^{\circ}$ when the reference point shifts from a position before the neutrino production point to one after it (see Fig. \ref{fig:theta_phi}). By monitoring the migration of the average $\varphi$ distribution, we can evaluate the distribution of neutrino production points along the decay volume. 


\subsection{Experimental observables with true neutrino direction }

Let us assume that we can reconstruct the neutrino direction ($\vec \nu$) with infinite precision. Consider a reference decay point $\vec r_0$, which lies along the decay volume: 
\begin{equation}
    \vec r_0^i = \vec r_{tgt} + \lambda\, \hat d_{dcy}, 
\end{equation}
where $\lambda=0$ corresponds to the target coordinate, and $\lambda = 110 $~m corresponds to the end of the decay volume. The values of $\lambda$ can be extended beyond these limits. We can construct the reference system as described in the previous section.

The angle $\varphi$ can be computed as:  
\begin{equation}
    \varphi = atan2( \vec \nu \cdot \vec y ,\vec \nu \cdot \vec x ).
\end{equation}

This calculation can be repeated for different reference points along the decay volume, leading to the definition of multiple reference frames and the examination of $\varphi$ angle distributions projected into each frame. When the reference point $\vec r_0$ coincides with the true neutrino origin, the $\varphi$ angle is undefined. For neutrino origins below this point (red arrow in Fig. \ref{fig:theta_phi}), the \(\varphi\) angle collapses to a value determined by the geometry of the experiment. For origins above the reference point (blue arrow in Fig. \ref{fig:theta_phi}), the \(\varphi\) angle collapses to a value that is rotated by 180$^{\circ}$ relative to the previous value. This behavior is shown for the beam and detector location similar to those in T2K. Based on the definition of the reference system (see section ~\ref{sec:reconstructed_neutrino_direction}), the expectation is that the $\varphi$ angle is -90$^{\circ}$ for neutrinos produced before the reference point along the decay volume, and +90$^{\circ}$ for those produced after this point. These angles are defined with respect to the $X$ axis of the reference system shown in Fig.~\ref{fig:theta_phi}.
The value of $\varphi$ as a function of the reference points is represented by using the Heaviside function ($\theta(x)$):

{\begin{equation}
    \varphi(x, x_0) = 90^\circ \left(1 - 2\,\theta(x - x_0)\right),
\end{equation} 
where $x$ is the reference point and $x_0$ is the neutrino origin. This function describes how $\varphi$ changes depending on whether $x$ is above or below $x_0$.

\begin{figure}
    \centering
    \includegraphics[width=0.75\linewidth]{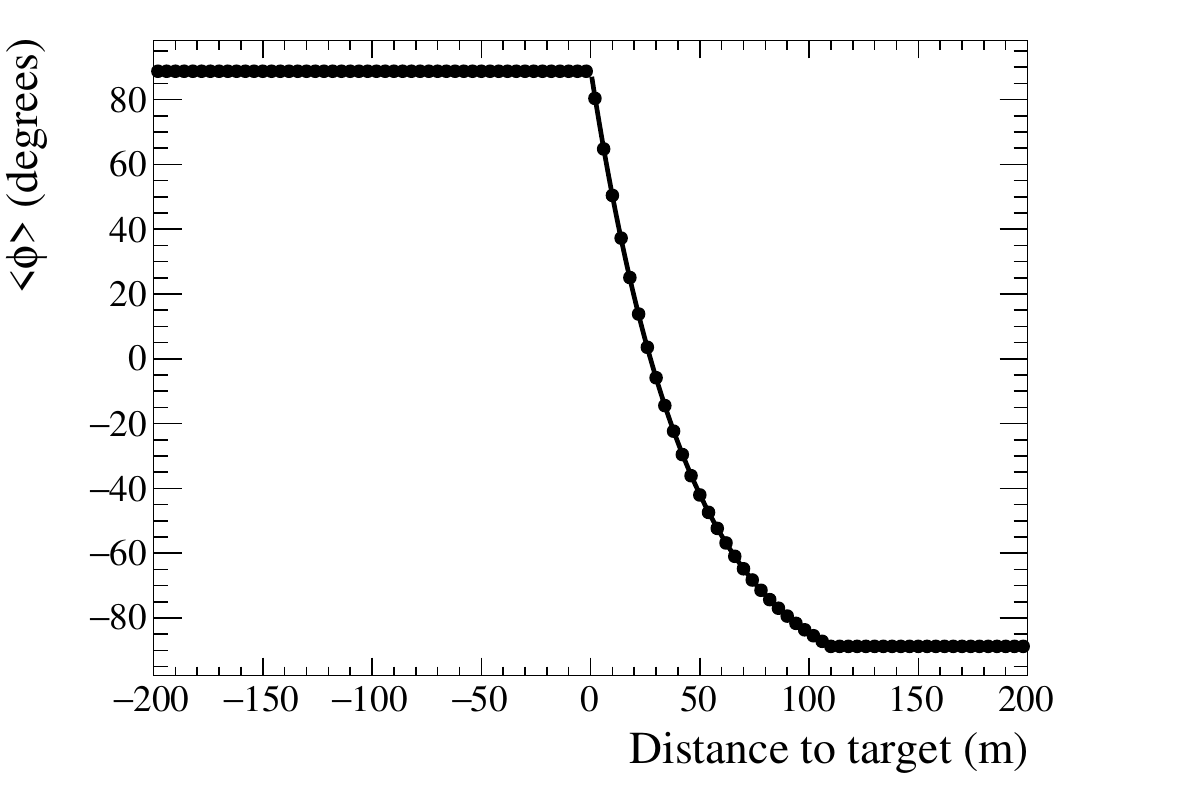}
    \caption{The true $\langle\varphi\rangle$$(x)$ as a function of the reference point for the exponential distribution of neutrino production points. The transition region shows an almost exponential dependence with a value close to $\lambda$.}
    \label{fig:TheorPhi}
\end{figure}

For a set of neutrinos originating from different points, we obtain a sum of step distributions that transition from $-90^\circ$ to $+90^\circ$. The average of these distributions is given by: 
\begin{equation}
    \langle \varphi(x) \rangle = \int p(x_0) \varphi(x, x_0) \, dx_0,
\end{equation}
where $p(x_0)$ is the probability density function of the neutrino production coordinate. Assuming that $p(x_0)$ follows an exponential distribution: 
\begin{equation}
    p(x_0) = \frac{1}{\lambda} e^{-x_0 / \lambda},
\end{equation}
we consider neutrinos produced between the target and the end of the decay volume at $110$~m. Although various values of $\lambda$ are discussed later in the paper, here we illustrate the distribution for $\lambda = 45$~m, resulting in the smooth transition shown in Fig. \ref{fig:TheorPhi}, assuming perfectly known neutrino directions produced along the beam axis.

The point where the average $\varphi$ angle transitions from positive to negative closely corresponds to the median position of neutrino production. In Fig.~\ref{fig:TheorPhi}, this transition occurs at 27.43~m, in excellent agreement with the median of the exponential distribution within the decay volume, calculated as 27.45~m. Under these conditions, the location where $\langle \varphi(x) \rangle$ crosses $0^{\circ}$ provides insight into the spatial distribution of neutrino production, effectively capturing the disintegration profile of the parent particles.

This result also reveals that the edge values of $\langle\varphi\rangle$ encode information about the detector’s position relative to the decay volume. When the reference point lies far upstream or downstream, the vector from this point to the interaction vertex (inside the detector) varies little across events, resulting in a nearly fixed local coordinate system. As a result, reconstructed neutrino directions cluster around a common orientation, and the azimuthal angle converges to $\pm90^\circ$, reflecting the stable geometric configuration. Deviations from these values thus signal a shift in the detector’s position relative to the decay volume, with any offset indicating a misalignment. We extended the exploration region to $[-200, 200]$~m to account for this geometric effect which becomes increasingly relevant farther from the decay volume. 

In the idealized case, using a Heaviside function and the true $\langle\varphi\rangle(x)$ distribution, assuming an exponential distribution of neutrino production points, this extension has no impact. 
However, as we will show later, the situation changes when the $\varphi$ angle is reconstructed with limited resolution. In that case, the sharp step-like behavior becomes smeared, producing a smooth transition in the circular mean and an asymptotic trend rather than a strict step function. Extending the region to large distances allows us to capture this edge behavior, which contains valuable information about the relative positioning of the detector and the decay volume.

\subsection{Experimental observables with reconstructed neutrino direction }

The use of the reconstructed neutrino direction introduces uncertainties in the calculation of the $\varphi$ angle. There are three primary sources of these uncertainties: 
\begin{itemize}
    \item Presence of the Fermi momentum (the motion of nucleons inside the nucleus).
    \item Presence of residual non-CCQE background.
    \item Detector effects: limitations in the detector’s resolution and efficiency can introduce additional distortions.
\end{itemize}

\begin{figure*}
    \centering
    \includegraphics[width=\linewidth]{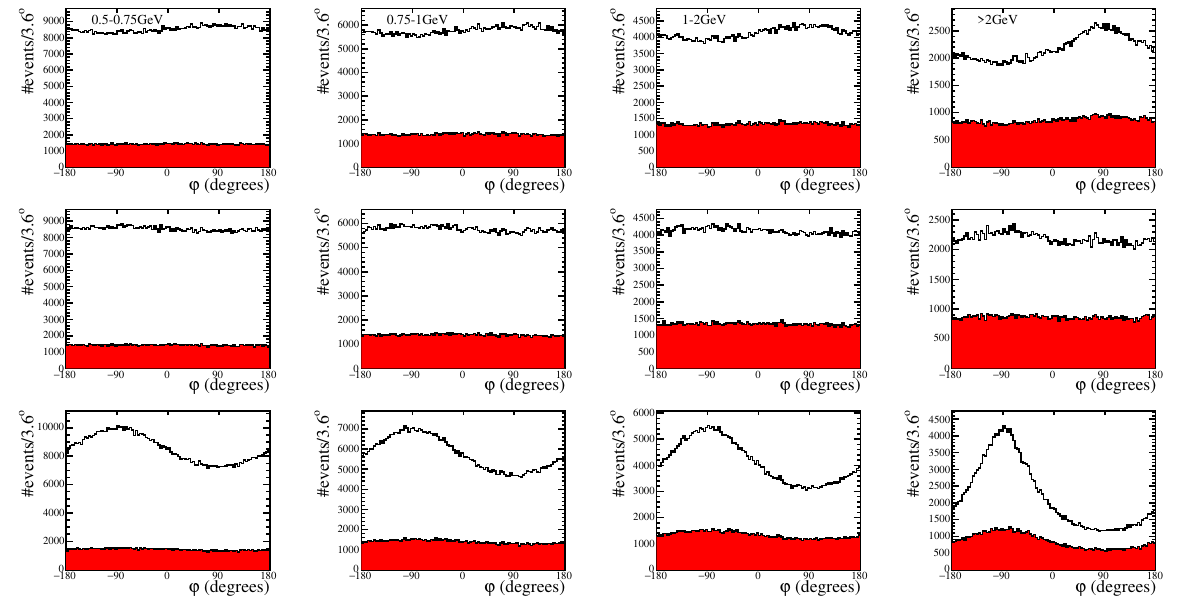}
    \caption{The distribution of $\varphi$ for four different energy regions (columns from left to right corresponding to increasing energy) and reference positions along the decay volume (rows from top to bottom representing the beginning, middle, and end of the decay volume). Events shown in red are non-CCQE contributions. The value of $\lambda$ used for these plots is fixed to 45~m.} 
    \label{fig:reconphi}
\end{figure*}

There are indications ~\cite{douqa2024superscaling} that the reconstruction of the neutrino direction using the muon and the leading proton ejected in the interaction preserves information about the neutrino direction, as discussed in the previous section. Using this definition of the neutrino direction, the $\varphi$ angle can be computed as:
\begin{equation}
    \varphi = atan2(\vec{p}_{\nu} \cdot \vec{y}, \vec{p}_{\nu} \cdot \vec{x})
\end{equation}
for different frames of reference. The results are shown in Fig. \ref{fig:reconphi} for different energy regions, with the reference point at the neutrino target (first row), the middle of the decay volume (middle row), and the end of the decay volume (last row). Results are shown for all exclusive events with muons and protons in the final state. We also identify those events that are not produced via CCQE interactions, shown in red in Fig. \ref{fig:reconphi}.

This is essentially a frame-induced asymmetry, fundamentally arising from the choice of reference frame. 
When the reference origin coincides with the true neutrino production point and the $Z$ axis is perfectly aligned with the true neutrino direction, the azimuthal angle $\varphi$ becomes undefined due to the absence of a transverse component (i.e. $x$ and $y$ equal $0$), its projection onto the transverse plane vanishes.
In our study, however, we consider the reconstructed neutrino direction, which differs slightly from the true direction due to smearing effects from neutrino-nucleus interactions, the reconstruction algorithm, and potential detector limitations (the latter not included in this analysis). These effects introduce a small transverse component, making $\varphi$ mathematically well-defined. 
Yet, because the reconstructed direction remains on average equal to the true one, the resulting $\varphi$ distribution remains flat, showing no angular preference.
When the reference point is displaced from the true neutrino production coordinates, the distribution of the $\varphi$ angle ceases to be uniform and displays characteristic angular modulations, as illustrated in Fig.~\ref{fig:reconphi}. Quantifying this angular dependence provides a means to investigate the spatial distribution of neutrino production points along the decay volume.

Although detector limitations are not explicitly accounted for in this analysis, the detector typically reconstructs particle directions with high precision. In this case, the transverse momentum is primarily determined by the reconstructed muon momentum, which generally exhibits excellent performance. Nuclear effects tend to dominate over detector resolution, ensuring that the main conclusions in this paper regarding the smearing behavior remain robust.

Attention to the Fig.~\ref{fig:reconphi} highlights several significant observations: 
\begin{itemize}
    \item \textbf{Shift in peak position:} The peak position of $\varphi$ shifts from positive to negative values as the reference point moves from the target to the end of the decay volume.
    \item \textbf{Peak width:} The width of the peak decreases, improving the sensitivity as the neutrino energy increases.
    \item \textbf{Resolution improvement:} The resolution of $\varphi$ improves as the reference point gets closer to the detector (last row). In this case, the polar angle $\theta$ is larger, and the resolution in $\varphi$ also improves. Conversely, as the distance increases (first row), the resolution worsens.
    \item \textbf{Behavior of background:} The background, shown as the red histogram in Fig. \ref{fig:reconphi}, does not show a dependence on the reference point except for high-energy neutrinos when the reference point is close to the end of the decay volume. This is a consequence of the previous two points.
\end{itemize}

These results indicate that the reconstructed $\varphi$ as a function of the location of the reference points can be used to estimate the distribution of neutrino production points along the decay volume and to evaluate the alignment between the neutrino beam and the detector.



\section{Extracting beam information from observables }
\label{sec:results}

We will explore several experimental observables based on the $\varphi$ angle. These include the average azimuthal angle ($\langle \varphi \rangle$), as well as trigonometric functions such as $\sin \varphi$ and $\cos \varphi$, along with the circular mean. These observables provide a suitable analytical framework for extracting relevant information from the data. For example, some observables are more effective at the edges of the decay volume, where geometric constraints or detector acceptance may amplify certain asymmetries or angular correlations. In contrast, other observables are more reliable in the central region of the decay volume, where statistical precision is higher and systematic effects due to boundary conditions are minimized.


\subsection{The $\langle \varphi \rangle$ observable }

The $\langle \varphi \rangle$ is the first observable that can be derived. Its advantage lies in its simplicity, as it is easy to treat statistically and directly represents the expected value of $\varphi$. However, a potential issue arises when background events are incorporated into the sample. In this context, we can define background as any event that does not exhibit a significant dependence on $\varphi$. This may occur either because particles are lost in the final state of the neutrino interaction, leading to poor reconstruction of the direction, or because the events are biased by the initial Fermi momentum of the target nucleon. In the presence of such background, the average value of $\varphi$ is typically shifted towards 0. This can be expressed as: 
\begin{equation}
\langle \varphi \rangle = f_S \langle \varphi \rangle_S + f_B \langle \varphi \rangle_B,
\end{equation}
where $f_S$ ($f_B$) are the fractions of the signal (background) and $\langle \varphi \rangle_S$ ($\langle \varphi \rangle_B$) are the average $\varphi$ for the signal (background). If we assume the average $\varphi$ for the background is 0 \footnote{This assumption may not hold in all cases, as the reconstructed direction could still carry information from the initial neutrino direction in the background.}, we obtain: 
\begin{equation}
 \langle \varphi \rangle = f_S \langle \varphi \rangle_S  < \langle \varphi \rangle_S.
\end{equation}

The $\langle \varphi \rangle$ as a function of the location of the reference points is presented separately for four energy regions and is shown in Fig. \ref{fig:Meanphi}. Each sub plot contains the results for the nominal position of the detector with respect to the decay volume and those with the detector shifted along the X or Y axes relative to the nominal position, which will be discussed later.

\begin{figure}[h]
    \centering
    \includegraphics[width=1.0\linewidth]{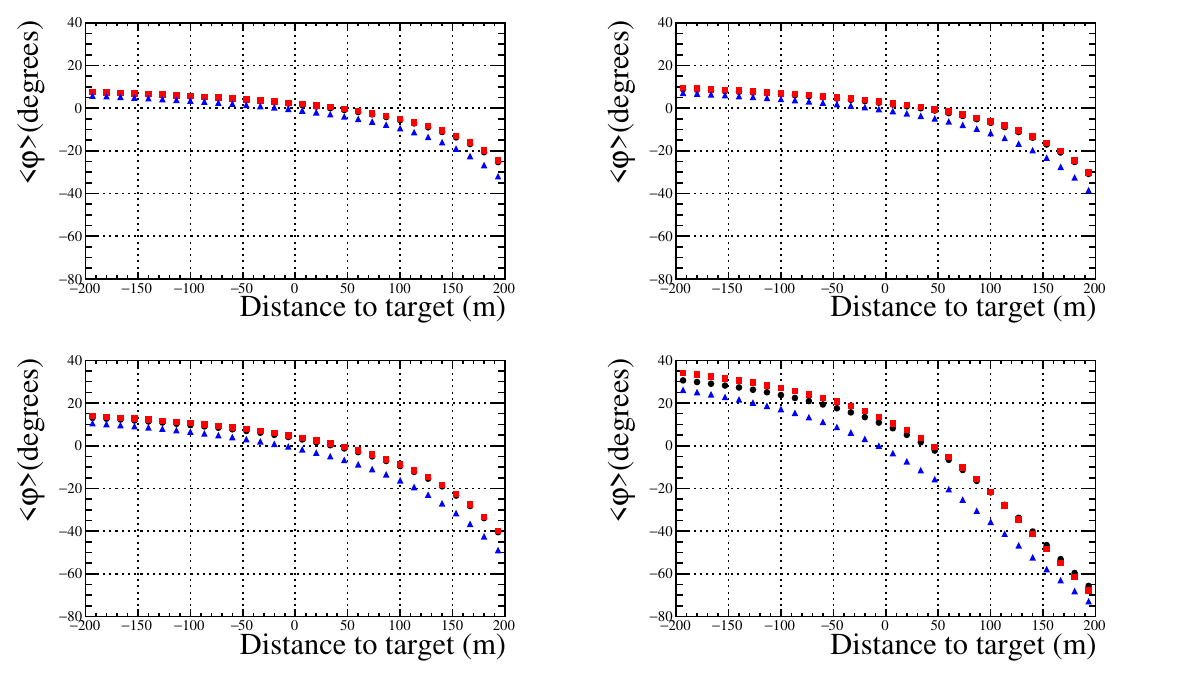}
    \caption{The $\langle \varphi \rangle $ as a function of the location of the reference points for $\lambda = 45$~m. The figures show the results for the nominal position (in black) and those for the fake detector misalignment of +2~m in X (in red) and +2~m in Y (in blue), presented separately for four energy regions :   
    0.5–0.75~GeV (upper left), 0.75–1~GeV (upper right), 1-2~GeV (lower left), $>$2~GeV (lower right). Notice that black and red graphs overlap in several sections of the plots. 
    }
    \label{fig:Meanphi}
\end{figure}

We calculate the distributions of $\langle \varphi \rangle$ as a function of the reference point (see Fig.~\ref{fig:Meanphi}) for various values of $\lambda$, and identify the corresponding positions at which $\langle \varphi \rangle$ equals 0$^\circ$. The resulting values are summarized in Table~\ref{tab:AvgPoint} for four distinct neutrino energy intervals ([0.5,0.75]~GeV,[0.75-1.0]~GeV,[1.0,2.0]~GeV and $>$ 2 GeV). It can be observed that the crossing values are larger than the expected mean (average) and median of the neutrino production point distribution. This is a bias introduced by errors in the reconstruction of the neutrino direction caused by the neutrino interaction nuclear effects. The $\theta$ angles for neutrinos in the reference frame are smaller for neutrinos before the reference point, leading to larger errors in $\varphi$ and potentially a slower transition to the edge value of -90$^{\circ}$.

Since $\pi$ and $K$ have different decay points for the same energy, we anticipate that a full numerical simulation of the experiment would enable exploration of their relative contributions. $K$ typically contribute to the high-energy neutrino component, where the position at which $\langle \varphi \rangle = 0$$^{\circ}$ reflects the particle distribution in the beam characterized by $\lambda$. More precise calculations will require comprehensive beam and detector simulations, which are beyond the scope of this work.

\begin{table}[h]
    \centering
    \begin{tabular}{c|c|c|cccc}
       & \multicolumn{2}{c|}{\textbf{True Value (m)}} & \multicolumn{4}{c}{\textbf{Position (m) at which } \textbf{$\langle \varphi \rangle = 0^\circ$}} \\
      \textbf{$\lambda$ (m)} &  \textbf{Mean} & \textbf{Median} & \textbf{0.5-0.75} & \textbf{0.75-1.0} & \textbf{1.0-2.0} & \textbf{$>$~2.0} \\
         \hline
      30  & 28.27 & 20.42 & 29.25 & 27.58 & 28.04 & 31.53 \\
      35  & 31.75 & 23.43 & 33.57 & 31.21 & 31.33 & 35.22 \\
      40  & 34.75 & 26.23 & 36.85 & 34.72 & 35.32 & 38.71 \\
      45  & 37.34 & 28.78 & 39.48 & 38.05 & 37.66 & 40.97 \\
    \end{tabular}
    \caption{Reconstructed average position (in meters) using the crossing point where $\langle \varphi \rangle$ = 0$^{\circ}$, for different neutrino energies and average true positions, obtained with exponential distributions of varying $\lambda$ values and for four energy regions (in GeV).}
    \label{tab:AvgPoint}
\end{table}

The reason for the higher crossing values lies in the difference between the median and the average in this context. The median represents the middle value of the neutrino origins, with half of the neutrinos originating before this point and half after. However, the average is influenced by the reconstructed angle $\varphi$. Neutrinos produced farther from the reference point tend to have their $\varphi$ angle more accurately reconstructed. Consequently, these more distant neutrinos have a greater impact on the average, shifting it away from the median concept and closer to that of the mean.

Furthermore, the geometrical model introduces an asymmetry: as we approach the end of the decay volume, the polar angle relative to the detector reference system becomes larger. This effect is evident in Fig. \ref{fig:reconphi}, where events reconstructed with the reference point placed at the end of the decay volume (last row) show a sharper peak compared to those with the reference near the target station (first row). This asymmetry results in a bias towards higher crossing values beyond the mean, which is also visible in Table \ref{tab:AvgPoint}. The specific bias will depend strongly on the detector's position relative to the decay volume, as well as the length of the decay volume and the distribution of neutrino production points. Therefore, it is important to predict this bias using the appropriate MC event generator. 

As illustrated in Fig. \ref{fig:Meanphi}, the slopes of the $\langle \varphi \rangle$ vary with neutrino energy. This variation indicates an improved determination of the $\varphi$ angle, as shown in Fig. \ref{fig:reconphi}. The angle also provides insights into the speed at which events migrate from positive to negative $\varphi$. To investigate these potential dependencies, we calculated the derivative of $\langle \varphi \rangle$ with respect to the distance to the target at the point where $\langle \varphi \rangle$=0$^{\circ}$. The results are presented in Table \ref{tab:Slope} for various neutrino energies and production point distributions within the decay volume. The dependency is very weak, but becomes more noticeable at low energies, primarily due to the large smearing induced by the poor $\varphi$ reconstruction at low energies. This smearing mixes larger regions of the decay volume in the calculations and makes the results more sensitive to the distribution of neutrino production points along the decay volume.

\begin{table}[h]
    \centering
    \begin{tabular}{c|cccc}
       & \multicolumn{4}{c}{\textbf{$d \langle\varphi\rangle/d \lambda$ at $\langle \varphi \rangle = 0$$^{\circ}$}} \\
      \textbf{$\lambda$(m)} & \textbf{0.5-0.75} & \textbf{0.75-1.0} & \textbf{1.0-2.0} & \textbf{$>$~2.0} \\ 
         \hline 
      30  & -0.062 & -0.076 & -0.11 & -0.28 \\
      35  & -0.064 & -0.078 & -0.11 & -0.28 \\
      40  & -0.065 & -0.080 & -0.12 & -0.29 \\
      45  & -0.067 & -0.081 & -0.12 & -0.29 \\
    \end{tabular}
    \caption{Derivative of $\langle \varphi \rangle$ with respect to the position at the crossing point, for different neutrino energies and average true positions, obtained using exponential distributions of varying $\lambda$ values.}
    \label{tab:Slope}
\end{table}



To investigate the sensitivity of $\varphi$ to the alignment of the detector with the beam, we introduced artificial displacements in the detector's $X$ and $Y$ coordinates. We then determined the values at which $\langle \varphi \rangle = 0^\circ$, as presented in Table \ref{tab:AvgPointShift}. A significant difference is observed between the shifts in $Y$ and $X$. This disparity arises from the specific relative positioning of the beam and detector. In our model, shifting the detector along the $Y$ axis keeps it within the plane containing the target and the decay volume. Conversely, a shift along the $X$ axis moves the detector out of this plane.

When the detector shifts positively along the $Y$ axis, the neutrino interaction coordinates move upward, while the true neutrino direction does not change in the detector. In this specific case, where the detector does not move out of the beam plane, neutrinos from a displaced interaction intersect the decay beamline closer to the target, resulting in a larger $Y$ coordinate. The exact $\varphi$ shift is complex and depends on the neutrino production point. However, generally, for any reference point between the two intersections (true and misaligned), the $\varphi$ angle changes from negative to positive values, shifting the crossing point, where $\langle \varphi \rangle = 0$$^{\circ}$, toward the negative direction. As shown in Fig. \ref{fig:Meanphi}, for neutrinos with $E_{\nu} > 2$ GeV, the extreme values are $26^{\circ}$ and $-72^{\circ}$ with $Y$ misalignment, compared to $30^{\circ}$ and $-65^{\circ}$ for the nominal position.  The average of the extreme values shows a shift in the $\langle \varphi \rangle$ from $(26^{\circ}-72^{\circ})/2 = -23^{\circ}$ to $(30^{\circ}-65^{\circ})/2 = -17.5^{\circ}$.

When the detector shifts positively along the $X$ axis, it moves out of the beam plane. In this scenario, the crossing point with the beam is minimally affected by the shift, but the absolute angles increase due to the perceived additional distance. Since the angle change varies for neutrinos near the target versus those near the end of the decay volume, we expect a shift in the value, though close to the nominal one. As illustrated in Fig. \ref{fig:Meanphi}, for neutrinos with $E_{\nu} > 2$ GeV, the extreme values for the $X$ misalignment are $34^{\circ}$ and $-67^{\circ}$, which are closer to the nominal values given in the previous paragraph. This indicates a tendency towards more positive crossing points as can be seen from the shift of the average of extreme values $(30^{\circ}-65^{\circ})/2 = -16.5^{\circ}$.

These examples demonstrates the potential of the $\langle \varphi \rangle$ observable. However, a more thorough analysis is required, tailored to the specific geometry of the experiment.

\begin{table}[h]
    \centering
    \begin{tabular}{c|c|cc}
        \textbf{Energy}      &  \multicolumn{3}{c}{\textbf{Crossing point (m)}} \\
                             & \textbf{Nominal} & \textbf{+2~m in X} & \textbf{+2~m in Y} \\
        \hline
         0.5-0.75 & 39.48 & 45.96 & -3.48 \\
         0.75-1.0 & 38.05 & 45.34 & -4.41 \\
         1.0-2.0 &  37.66 & 44.71 & -6.22 \\
         $>$~2.0  & 40.97 & 46.91 & -3.93 \\ 
         \end{tabular}
    \caption{ Crossing point where $\langle \varphi \rangle = 0$$^{\circ}$ for $\lambda = 45$~m with a 2~m misalignment introduced in the detector’s X and Y coordinates. }
    \label{tab:AvgPointShift}
\end{table}

\subsection{The $\cos{\varphi}$ and $\sin{\varphi}$ }

The functions $\cos{\varphi}$ and $\sin{\varphi}$ can be seen as the first components of the Fourier transform of the $\varphi$ distribution. They have the advantage of extracting more information from the distribution, as $\langle \sin \varphi \rangle$ and $\langle \cos \varphi \rangle$ are independent measurements when sufficient statistics are available. $\langle \sin \varphi \rangle$ and $\langle \cos \varphi \rangle$ contain information about the rotation between the decay volume and the detector and can facilitate the relative rotation of both references. In our case study, the detector is located in the plane containing both the decay volume and the target position, so $\langle \cos \varphi \rangle$ is expected to be 0 (see Fig. \ref{fig:AvgCos}). The result with a large displacement along the $X$ coordinate (in red in Fig. \ref{fig:AvgCos}) shows a large impact on $\langle \cos \varphi \rangle$ from the detector to the beam alignment.

$\langle \sin \varphi \rangle$ shows a more pronounced dependency similar to that of $\langle \varphi \rangle$ (see Fig. \ref{fig:AvgSin}). In this case, the main misalignment is induced by the detector displacement along $Y$ (in blue in Fig. \ref{fig:AvgSin}). Like in the case of $\langle \cos \varphi \rangle$, the larger energy regions show stronger variations.

\begin{figure}[h]
    \centering
    \includegraphics[width=1.0\linewidth]{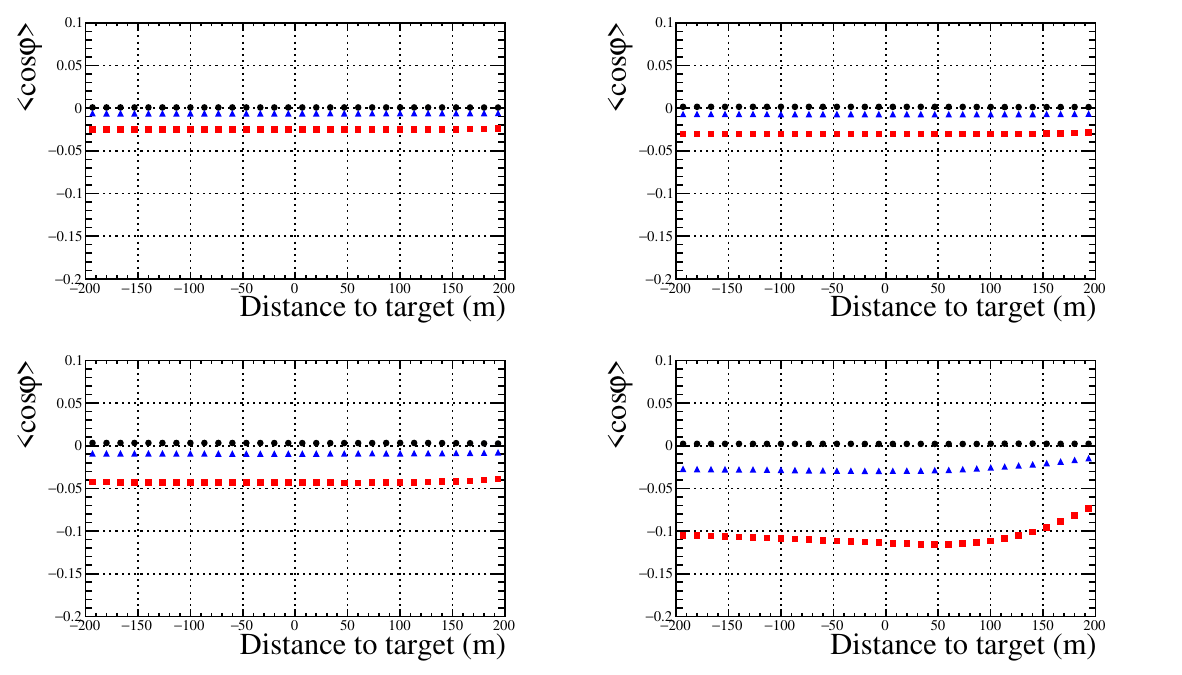}
    \caption{$\langle \cos \varphi \rangle$ as a function of the location of the reference points for $\lambda = 45$~m. The figures show the results for the nominal position (in black) and those for the fake detector misalignment of +2 m in X (in red) and +2 m in Y (in blue) separately for four energy regions :   
    0.5–0.75~GeV (upper left), 0.75–1.0~GeV (upper right), 1.0-2.0~GeV (lower left), $>$~2.0~GeV (lower right).} 
    \label{fig:AvgCos}
\end{figure} 

\begin{figure}[h]
    \centering
    \includegraphics[width=1.0\linewidth]{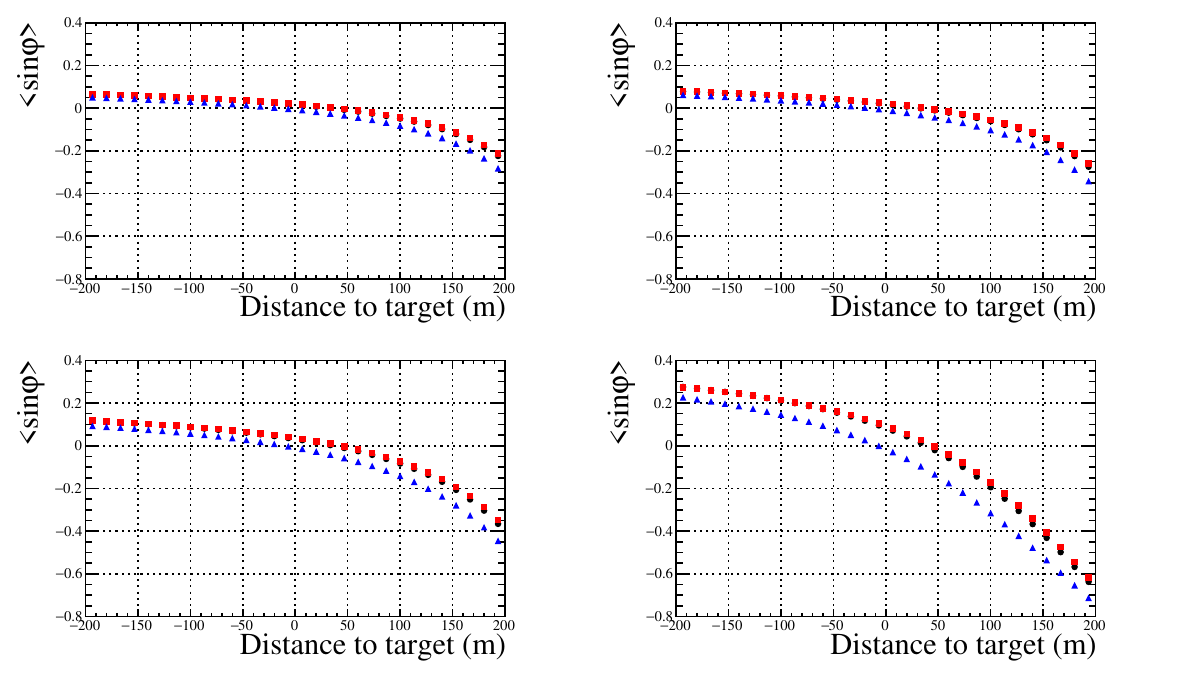}
    \caption{$\langle \sin \varphi \rangle$ as a function of the location of the reference points for $\lambda = 45$~m. The figures show the results for the nominal position (in black) and those for the fake detector misalignment of +2 m in X (in red) and +2 m in Y (in blue), presented separately for four energy regions:   
    0.5–0.75~GeV (upper left), 0.75–1.0~GeV (upper right), 1.0-2.0~GeV (lower left), $>$~2.0~GeV (lower right). Notice that black and red graphs overlap in several sections of the plots.} 
    \label{fig:AvgSin}
\end{figure} 

Higher-order $\varphi$ dependencies, such as $\sin{k \varphi}$ and $\cos{k \varphi}$ with $k > 1$, can be explored. However, based on the discussion above and as shown in Fig. \ref{fig:reconphi}, we expect the distributions to exhibit a single-cycle oscillation ($k=1$).


\subsection{The circular mean }


The circular mean $\varphi_C$ is defined as: 
\begin{equation}
 \varphi_C = atan2( \langle \sin \varphi \rangle, \langle \cos \varphi \rangle ) 
\end{equation}
This definition overcomes the issue mentioned above, as the factor $f_S$ 
\begin{equation}
 \langle \sin \varphi \rangle = f_S \langle \sin \varphi \rangle_S
\end{equation}
\vspace{-0.8cm}
\begin{equation}
 \langle \cos \varphi \rangle = f_S \langle \cos \varphi \rangle_S  
\end{equation}
cancels in the arc-tangent calculation. This comes with a disadvantage. The transition region ($\varphi_C\approx 0$$^{\circ}$) is not well defined and has large uncertainties. For example, if the value of $\langle \cos \varphi \rangle$ is close to 0, its experimental value can fluctuate between positive and negative, leading to transitions between a 0$^\circ$ crossing (when $\langle \cos \varphi \rangle$ is positive) and a 180$^{\circ}$ crossing (when $\langle \cos \varphi \rangle$ is negative). As a result, the uncertainty in this region is large, making estimation challenging. In contrast, the values at the edges are expected to converge to the theoretical predictions derived in the previous sections.

The results are shown in Fig. \ref{fig:CircularMean}, where they are fitted to a sigmoid curve, and the crossing point is determined, as shown in Table \ref{tab:CircularMean}. These results are comparable to those of Table \ref{tab:AvgPoint}, which is not surprising, as they are dominated mainly by the crossing of $\langle \sin{\varphi} \rangle = 0$ (see Fig. \ref{fig:AvgCos}). 

\begin{figure}[h]
    \centering
    \includegraphics[width=1.0\linewidth]{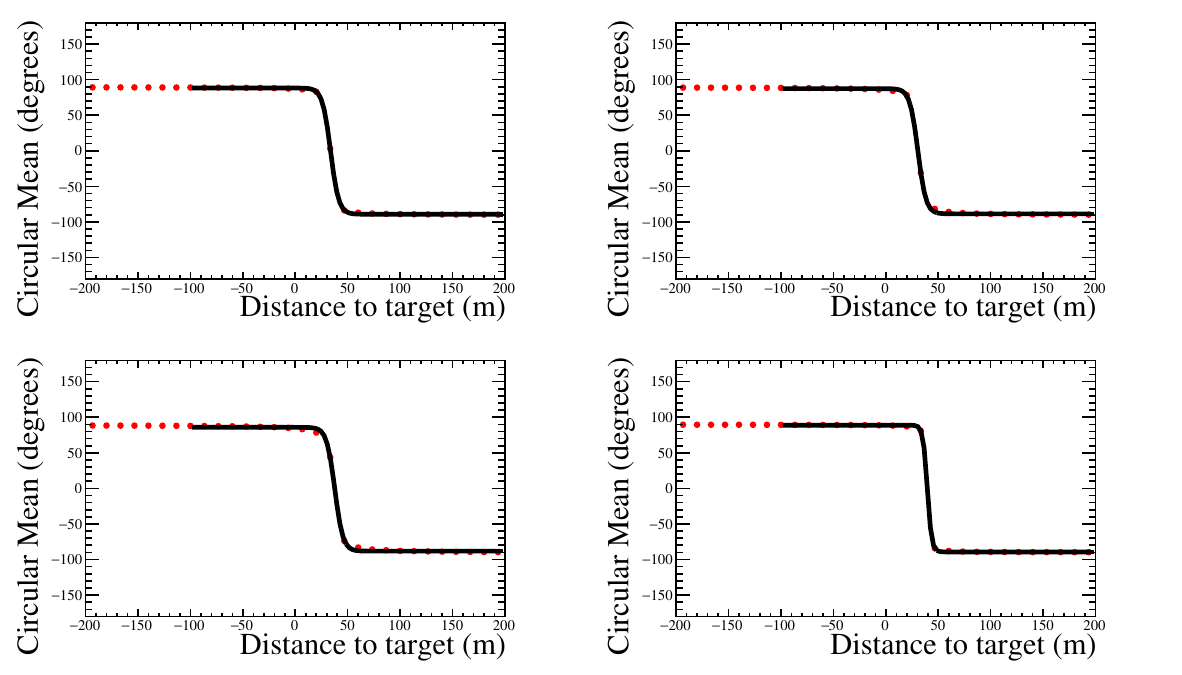}
    \caption{The circular mean as a function of the location of the reference points for $\lambda = 45$~m, presented separately for four energy regions:   
    0.5–0.75~GeV (upper left), 0.75–1.0~GeV (upper right), 1.0-2.0~GeV (lower left), $>$~2.0~GeV (lower right). }
    \label{fig:CircularMean}
\end{figure}

The values at the edges are clearly -90$^{\circ}$ and +90$^{\circ}$, as expected, given the relative position of the detector with respect to the decay volume. This is actually a consequence of $\langle \cos{\varphi} \rangle$ being $\approx 0 $.
\begin{table}[H]
    \centering
    \begin{tabular}{c|c}
          \textbf{Energy bin (GeV)} & \textbf{Crossing point (m)}\\
         \hline
          0.5-0.75 & 40.025\\
         0.75-1.0 & 39.37 \\
         1.0-2.0 & 39.71 \\
         $>$~2.0 & 40.45 \\ 
    \end{tabular}
    \caption{Crossing point obtained with the circular mean for the case of $\lambda = 45$ (mean 37.34 and median = 28.78).}
    \label{tab:CircularMean}
\end{table}

\section{Conclusions and prospects }
\label{sec:conclusions}

We have demonstrated that the variation of $\varphi$, reconstructed using different reference points, is sensitive not only to the average neutrino production point in the decay volume, but also to the alignment and rotation of the detector relative to the decay volume. The variable $\varphi$ retains information about the neutrino production point by calculating its average. Other variables, such as $\cos{\varphi}$, $\sin{\varphi}$, and the circular mean, were also discussed, revealing different dependencies on beam parameters.  

The different $\varphi$-based observables enable us to statistically distinguish between neutrinos produced before or after the origin of the selected reference frame along the decay volume. They also facilitate cross-checking the MC simulations against the beam model. These observables provide valuable insights into detector-beam alignment, beam dynamics, such as the average decay position of parent particles in the decay volume, and other factors influencing the expected neutrino flux at the near detector.
For future neutrino experiments or upgraded detector techniques with higher statistical precision, this method will be essential for estimating the average decay position from experimental data and for identifying potential modeling of the neutrino beam.  

\begin{acknowledgments}
This work was supported by the Federal Commission for Scholarships for Foreign Students for the Swiss Government Excellence Scholarship (ESKAS No. 2022.0154) for the academic years 2022-2025 and the Swiss National Science Foundation (SNF) grant (No. 200020\_204609). The authors extend their gratitude to the T2K neutrino beam group and the University of Geneva neutrino group for their valuable discussions and insights. Additionally, we thank Prof. K. McFarland for his insightful discussions during the completion of the manuscript.
\end{acknowledgments}

\bibliography{apssamp}

\end{document}